\documentclass[article,nojss]{jss}
\usepackage[utf8]{inputenc}
\usepackage[T1]{fontenc}
\usepackage{fixltx2e}
\usepackage{graphicx}
\usepackage{longtable}
\usepackage{float}
\usepackage{wrapfig}
\usepackage{rotating}
\usepackage[normalem]{ulem}
\usepackage{amsmath}
\usepackage{textcomp}
\usepackage{marvosym}
\usepackage{wasysym}
\usepackage{amssymb}
\usepackage{hyperref}
\usepackage[draft]{fixme}
\fxusetheme{color}
\FXRegisterAuthor{gb}{agb}{GB}
\FXRegisterAuthor{ml}{aml}{ML}
\FXRegisterAuthor{rg}{arg}{RG}
\Address{ Gabriel Becker \\ Dept. of Bioinformatics and Computational Biology \\ Genentech Research and Early Development \\ 1 DNA Way, South San Francisco, CA, 94080 United States of America \\ Email: \email{becker.gabriel@gene.com}}
\Abstract{Science depends on collaboration, result reproduction, and the
development of supporting software tools.  Each of these requires
careful management of software versions.  We present a unified model
for installing, managing, and publishing software contexts in R.
It introduces the package manifest as a central data structure
for representing version-specific, decentralized package cohorts.
The manifest points to package sources on arbitrary
hosts and in various forms, including tarballs and directories under
version control. We provide a high-level interface for creating and
switching between side-by-side package libraries derived from
manifests.  Finally, we extend package installation to support the
retrieval of exact package versions as indicated by manifests, and to
maintain provenance for installed packages.  The provenance
information enables the user to publish libraries or sessions as
manifests, hence completing the loop between publication and deployment.
We have implemented this model across two software
packages, switchr and GRANbase, and have released the source code
under the Artistic 2.0 license.}
\author{Gabriel Becker, Cory Barr, Robert Gentleman, Michael Lawrence}
\date{\today}
\title{Enhancing reproducibility and collaboration via management of R package cohorts}
\hypersetup{
  pdfkeywords={},
  pdfsubject={},
  pdfcreator={Emacs 24.3.1 (Org mode 8.2.10)}}
\begin{document}

\maketitle

\section{Introduction}
\label{sec-1}

Every data analysis rests on four pillars: the data, the analysis
code, the statistical methods, and the software which implements the
methods. Changes to any one of these pillars will affect analysis
results. We focus our attention on the often overlooked fourth pillar:
the exact set of software --- including the specific versions thereof
--- used to perform an analysis. We call this the \emph{software context} of
the analysis.

Management of software contexts is relevant to many activities in
scientific computing.  Collaborators often need to synchronize package
versions to guarantee comparability of their results \citep{gentleman_statistical_2004}
\citep{rstudio_inc_packrat_2014} \citep{fitzjohn_reproducible_2014}. A package author
might switch between contexts when maintaining multiple software
branches, or when alternating between development and analysis work
\citep{ooms_possible_2013} \citep{gentleman_bioconductor_2003}.  Large
collaborative or enterprise organizations might formally support this
synchronization by automating the testing and publication of a
canonical cohort of package versions \citep{revolution_analytics_mran_2014}. 
Finally, analysts attempting to reproduce published computational results
often need to approximate the software context of the original
authors \citep{rstudio_inc_packrat_2014}.

The above examples suggest three general software requirements for
managing software contexts. First, users need to locally manage
software contexts, including the ability to create,
populate, and conveniently switch between them. Second,
specific package versions must be directly installable, including from
non-repository sources \citep{wickham_devtools_2014} and historical releases
\citep{revolution_analytics_mran_2014}. Finally, organizations and
individuals should be able to define, test, and publish specialized,
version-specific package cohorts \citep{rstudio_inc_packrat_2014},
\citep{revolution_analytics_mran_2014},
\citep{de_vries_minicran_2014}. Users should be able to install packages
from such a cohort through standard mechanisms, or use the cohort to
seed a new software context.  

\cite{rstudio_inc_packrat_2014}'s \textbf{packrat} package focuses on
packages installed and used by a particular analysis project. It
provides the ability to maintain separate package libraries for
different projects, and to bundle the package tarballs themselves with
a project's analysis script(s) for preservation and manual
distribution, resulting in what \cite{gentleman_statistical_2004} call \emph{compendiums}. \cite{revolution_analytics_mran_2014}'s approach with 
\textbf{MRAN}, on the other hand, focuses on preserving the state
of CRAN in its entirety at specific timepoints. These \emph{snapshots} are
preserved as live repositories which can be installed from at a
later date as necessary. Other programming languages have similar
tools, including Perl \citep{miyagawa_carton_2011},
\citep{trout_local-lib_2007} and Python \citep{bicking_virtualenv_2007},
though they typically focus on development environments rather than
collaboration or analysis reproducibility.

We present a framework for managing and distributing software contexts
for the \proglang{R} statistical computing language. The framework provides two
separate but integrated functionalities. First, we provide tools for locally
installing and managing software contexts, including the retrieval of exact,
specified package versions from a variety of sources. Secondly, we
provide tools for publishing (optionally validated)
version-specific R package cohorts which can be deployed
locally as self-contained software contexts.

\section{A brief example}
\label{sec-2}

Suppose we are embarking on a large-scale
collaboration with a group of other researchers. Recognizing the need
for comparable results, we all agree to standardize our software
contexts on the package versions available from CRAN and
Bioconductor on that date.

After installing or updating the relevant packages, we can create a package manifest
which describes our set of currently installed packages --- i.e., our
current \emph{package library}, not to be confused with the
\verb!library()! function ---  via the
\verb!libManifest()! function.

\begin{Code}
library(switchr)
pkg_man <- libManifest()
\end{Code}

This manifest will act as a generalized repository, allowing
collaborators to consistently install the same package versions across
time and physical distance. We could also construct a manifest
directly from the cohort of packages
available from one or more package repositories --- e.g., CRAN and 
Bioconductor --- or from a \textit{\textbf{SessionInfo}} object.
We discuss the details of how package
manifests provide the cornerstone of our framework in Section \ref{sec-3}. 

To ease our collaborators' use of the manifest, we can use the
\textbf{switchrGist} package to publish the manifest as a GitHub gist:

\begin{Code}
library(switchrGist)
gisturl <- publishManifest(pkg_man, Gist())
\end{Code}

Now suppose a collaborator wishes to install the chosen package
versions using our manifest. She can avoid overwriting currently
installed packages with those chosen for our collaboration by \emph{switching to} a new package library and \emph{seeding} it with the
package manifest (installation output omitted for brevity).

\begin{Code}
switchTo("CollabEnv", seed = pkg_man)
\end{Code}

\begin{verbatim}
Switched to the 'CollabEnv' computing environment. 43 packages are currently 
available. Packages installed in your site library ARE suppressed.
 To switch back to your previous environment type switchBack()
\end{verbatim}

Switching to a package library has three primary effects. First, it unloads
any currently loaded packages from the \proglang{R} session. Next, it resolves
the specified name into a library, creating a new one if no library
with that name exists. If a new library is being created and a \emph{seed}
is provided, the packages listed in the seed are automatically
installed during the creation process; seeds are ignored in the case
of a pre-existing library. Finally, it configures the current \proglang{R} session to
use the specified library. After switching, the collaborator can use
the specified packages without affecting her default --- or
any other --- library.

The collaborator would then proceed to work on the project by using packages within
the "CollabEnv" library.  To work on something else within the same
\proglang{R} session, she can \emph{switch back} to the previous library, which was unaffected by both the seeding
and any subsequent package installation:

\begin{Code}
switchBack()
\end{Code}

\begin{verbatim}
Reverted to the 'original' computing environment. 208 packages are currently 
available. Packages installed in your site library ARE NOT suppressed.
 To switch back to your previous environment type switchBack()
\end{verbatim}

When returning to working on the collaboration --- either in the same R
session or within a different one --- the collaborator simply switches
to her existing, specialized library:

\begin{Code}
switchTo("CollabEnv")
\end{Code}

Alternatively, our collaboration might decide that a shared and
(potentially) evolving set of package versions serves our purpose
better. In this case, we can use \textbf{GRANBase} to create a
traditional, validated package repository from a package manifest:

\begin{Code}
library(GRANBase)
makeRepo(manifest, baseDir = "~/collabrepos",
	 repoName = "CollabPkgs")
\end{Code}

\section{Methods}
\label{sec-3}
\subsection{Representing package cohorts via generalized repositories}
\label{sec-3-1}

Every \proglang{R} workflow involves specific cohorts of
packages. Installed packages are collected into package libraries, which
dictate the set of packages loadable within R. Loaded and attached
packages control the set of functionality available to the user within
his or her \proglang{R} session. The cohorts of packages which pass check together
define the contents of the CRAN and Bioconductor package
repositories, determining which packages --- and versions thereof ---
users can install via standard machinery.

Package repositories provide a natural, existing mechanism for
representing and publishing cohorts of packages beyond their narrow
--- but indispensible --- current role of defining the R software
ecosystem at large. A package repository is essentially a mapping
between a set of package names and one or more pre-built archives of
the package source code or binaries.  Using a package repository, we
can define and publish arbitrary package cohorts, which end-users can
install via the standard package installation machinery in \proglang{R}.
For example, a cohort might correspond to the set of package versions
used to generate the results of a single publication. Broader
application of package repositories would enhance reproducibility,
package development and collaboration.

We generalize the concept of package repositories via
\emph{package manifests}. Package manifests define package cohorts ---
including the information necessary to retrieve and install the
packages' source code --- decoupled from the pre-built tarballs
stored in standard repositories. Instead, package sources can reside
in virtually any form, including directories under public (GitHub,
Bitbucket) or private source control, a package within a standard
repository or the CRAN archive, or more generally a Web accessible
directory or tarball. This decoupling allows users to create, install
from, and publish package cohorts --- including those that contain
packages or versions thereof which are not availablein standard
repositories --- without specialized hosting.

\begin{figure}[htb]
\centering
\includegraphics[width=.75\textwidth]{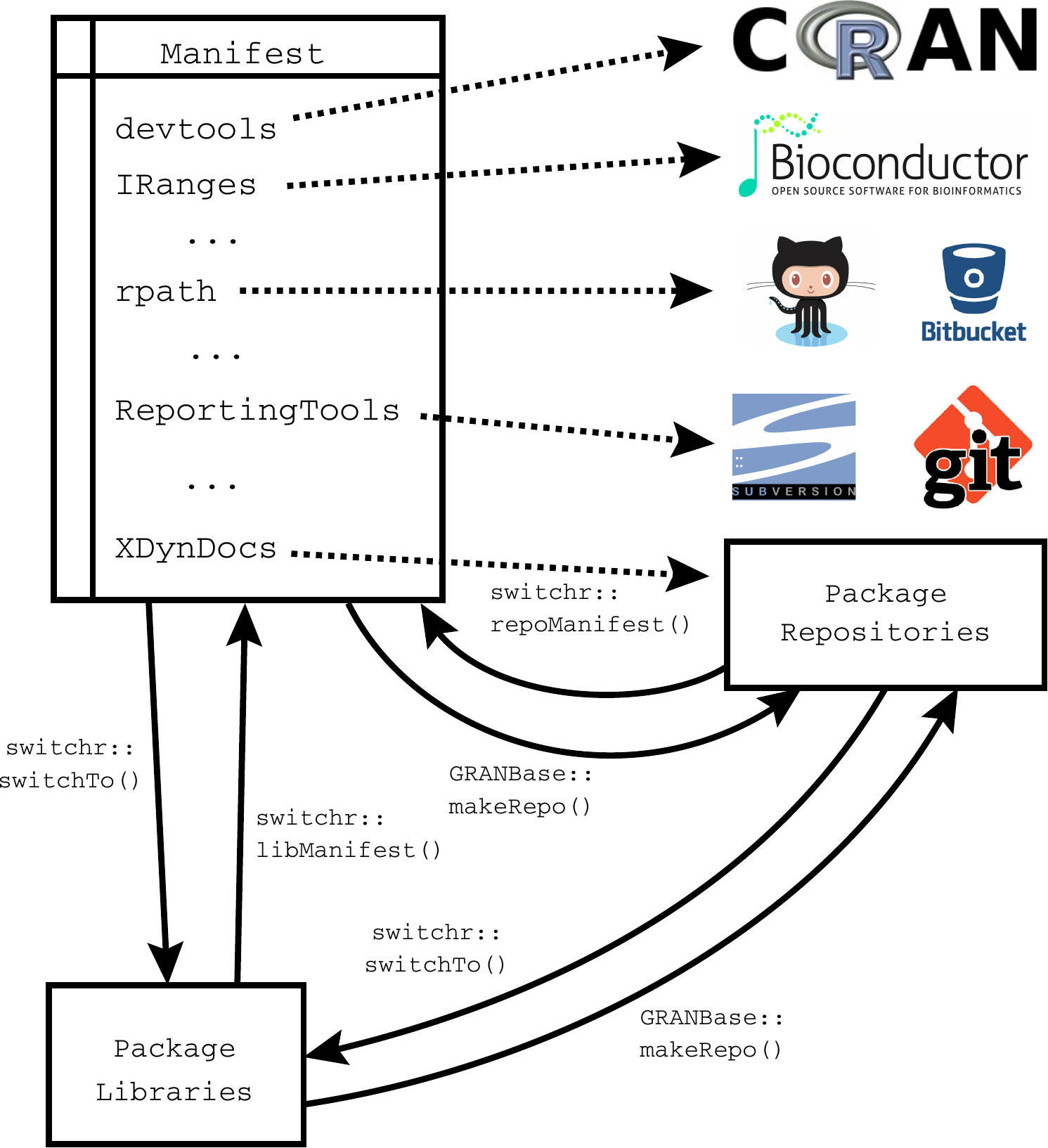}
\caption{\label{archPic}Our framework centers around representing package cohorts as manifests. These manifest can be transformed into useful forms, such as package libraries, and validated repositories.}
\end{figure}

Via the manifest abstraction, we can operate at the level of entire package
cohorts (as approximations of software contexts), rather than individual packages. Our framework allows users
to bi-directionally and reversibly transform package cohorts between three
forms: abstract manifests, installed package libraries, and standard
package repositories, as pictured in Figure \ref{archPic}. This allows users to
describe, share, publish, locally recreate, and use software contexts
directly (R version, OS, and external programs not withstanding).

A \emph{seeding manifest} represents a filtered subset of a package
manifest. Filtering allows users to define and install a subset of a
larger package manifest, such as one representing a community
repository. Currently supported filters are package, indicating
inclusion in the subset, and version, indicating an exact version of
the associated package. We use seeding manifests to represent both
package libraries --- the set of installed package versions ---  and
the set of packages loaded within the current \proglang{R} session, along
with the information necessary to reinstall that set of packages elsewhere.

\subsection{An abstraction for managing package libraries}
\label{sec-3-2}

Our \emph{switching} abstraction, presented briefly in Section \ref{sec-2},  combines all activities necessary for the user to begin
using the specified library --- whether or not it exists prior to the
switch. These include: altering the library path (the location on disk
where \proglang{R} installs and loads packages), unloading currently loaded
packages, and, if necessary, creating the library and \emph{seeding} it
with a set of packages. 

Seeding a package library, e.g., with a seeding manifest,
automatically installs the selected package versions into the library
during creation. This provides a convenient mechanism for recreating
the \proglang{R}-package portion of a published software context. In
principle, we can seed libraries with any object which specifies a set
of package versions along with their locations, though typically a
package or seeding manifest is used.

Alternatively, the user can derive a library from an existing one by
either \emph{branching} or \emph{inheritance}. If a library inherits from
another, the packages installed in the derived library override those
in the parent, while other packages in the parent remain
available. Branching, on the other hand, copies the installed contents
of an existing library into the new one but otherwise preserves no
relationship between them.

Updating package versions within a library mid-project can be costly,
potentially involving the regeneration of results or the modification
of analysis or package code \citep{ooms_possible_2013}. Freezing package versions over the course of
long-running projects, however, introduces a risk of generating incorrect
results \emph{which would have been correct if generated using up-to-date
software}.

Weighing the costs and benefits of updating against those of using
currently installed versions is an important aspect of library
management. The optimal strategy will vary greatly depending on the
needs of specific projects or users. We facilitate these decisions by
introducing the concept of an \emph{update risk report} which summarizes
the available updates and, to the extent possible, their potential impacts.

\subsection{Extending package libraries with provenance}
\label{sec-3-3}

We saw in Section \ref{sec-2} that a user can export a session or
library to a package manifest. Traditional package
libraries, however,  do not contain provenance information regarding
how --- and from where --- the package was installed. Without this
information, construction of an accurate manifest is difficult.

Our extension of the base \proglang{R} installation mechanism, which is discussed
more generally in Section \ref{sec-3-5},
automatically records provenance information for packages as they are
installed, generalizing the approach taken by
\cite{wickham_devtools_2014} in \textbf{devtools}'s \verb!install_github()! function.  With package provenance recorded at installation time ---
supplemented via heuristics for packages installed via other
methods ---  we can transform a set of installed packages directly into a seeding (or
package) manifest. This allows full, bi-directional conversion between
the package library and package manifest representations of a given
cohort.

The user can publish an exported package manifest to the Web,
e.g., as a GitHub gist or a simple Web-hosted file, allowing others to
immediately recreate or install from the described cohort. This allows
users to effectively publish package libraries, ensuring others,
whether they be current collaborators or future researchers attempting
to reproduce our results, can recreate them. Package
cohorts can also be published in the form of a traditional, validated 
package repository which we discuss in more detail in Section \ref{sec-3-6}.

\subsection{Just-in-time repositories}
\label{sec-3-4}

Many package versions reside outside the package ecosystem defined by
a given set of repositories. For example, development package versions
might reside on GitHub, while superseded historical versions might be
found in the CRAN Archive or Bioconductor SVN, etc. Installing these can be challenging,
particularly in the face of dependencies between such packages. We
introduce  \emph{Just-in-time} (JIT) package repositories to mitigate these
challenges and allow direct, dependency-aware package installation in
the absence of a pre-existing repository.

JIT repositories are transient, local package
repositories, which are constructed at installation time and populated
with only the requested packages and their (potentially non-repository-available)
dependencies. Once the JIT repository is created, actual installation
can be delegated to R's established, core installation machinery, as
we do in the \verb!install_packages()! function in our \textbf{switchr}
package.

The JIT repository mechanism does not provide any CRAN- or
Bioconductor-like guarantees that a package will install on a
particular OS, or that a cohort of package versions are compatible. It
does, however, allow users to conveniently install cohorts of
inter-dependent packages as-is from a combination of repository and
non-repository sources.

\subsection{Extending R's package installation mechanism}
\label{sec-3-5}

Leveraging JIT repositories, \textbf{switchr} provides an
extension to \proglang{R}'s package installation mechanism with three key
features:
\begin{enumerate}
\item Installation of packages from package manifests (including
dependency support);
\item Installation of specific package versions --- including non-current
ones --- from most types of package location; and
\item Automatic recording of package provenance during installation.
\end{enumerate}

We provide a mechanism which supports installing packages --- or
specified versions thereof --- from package manifests, but which
utilizes \proglang{R}'s core, existing installation machinery (\verb!install.packages()!) under the hood. This
maintains a single code-path for installation logic internally while expanding
the options of end-users. 

When  installing specific package versions, \textbf{switchr} automatically searches
through a number of different locations. These include current and
historical CRAN releases, the package repositories for all
Bioconductor releases, and the commit history of SCM systems,
including the Bioconductor SVN and any listed in the relevant
entry in the package manifest. 

\begin{figure}[htb]
\centering
\includegraphics[width=.9\linewidth]{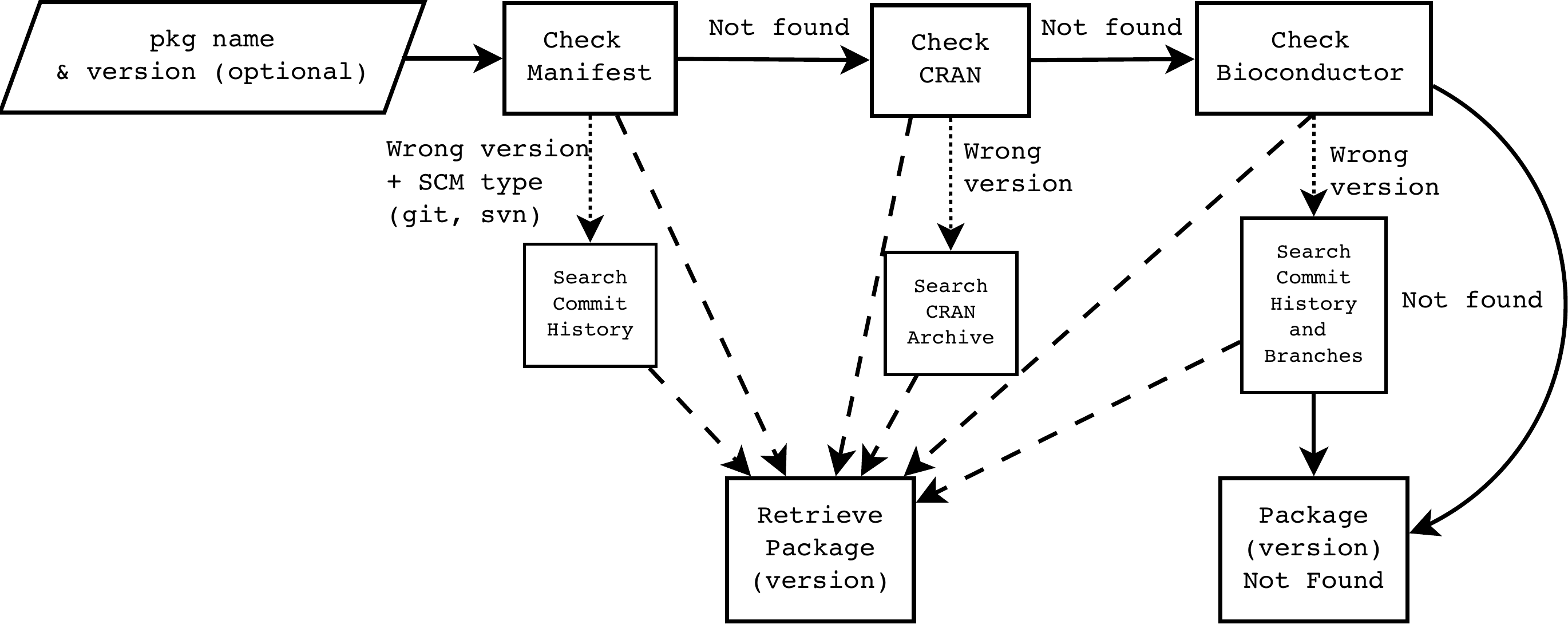}
\caption{\label{searchalgo}Package versions are searched for in three stages. First the manifest is checked, including SCM history for git and SVN package locations. Next, CRAN and the CRAN archive are searched. Finally, Bioconductor repositories and SVN histories are searched.}
\end{figure}

The specific search algorithm happens in three stages: searching the
manifest, searching CRAN, and searching Bioconductor, as pictured in
Figure \ref{searchalgo}.

When retrieving historical releases from SCM commit history, many
commits can define the same package version. To remove this ambiguity,
we define the commit associated with a package version number to be
the \emph{earliest} commit with that version listed in the package's
DESCRIPTION file. This models the process whereby developers might
make changes without changing the version number, but those changes
are not (ever) reflected in that version of the package.

\subsection{Building and testing package cohorts into validated repositories}
\label{sec-3-6}

Packages or package versions which appear together in a package or
seeding manifest are not, in general, guaranteed to be compatible with
each other. This is particularly true when the manifest points to the
SCM locations of development versions of multiple packages. For important cohorts of packages, it is beneficial that
they be built and tested together --- a service that traditional,
validated repositories such as CRAN and Bioconductor provide. 

We can construct validated package repositories from package
manifests. These repositories are tested \emph{incrementally}, meaning that when
testing a cohort of packages, only those packages which have been
updated since they last passed --- or have dependencies which have ---
are checked. A detailed build report  is
automatically generated and placed within the 
\verb!contrib! directory of the new repository.
 These features facilitate frequent, automated
testing via continuous integration systems such as Jenkins
\citep{kawaguchi_welcome_2014} or
Travis \citep{travis_ci_gmbh_travis_2014}. We note here that while our framework supports constructing an
actual repository, a manifest which simply points directly to the
pre-built tarballs which passed the integration testing would serve a
similar purpose and is also supported.

\section{Results}
\label{sec-4}
Here we present some example applications of our framework. These
include the recreation of historically published results, the
retrieval and installation of packages and their dependencies from
GitHub, and the publication of package manifests for use by the wider community.

\subsection{Recreating Anders and Huber's DESeq paper four years later}
\label{sec-4-1}
We now apply \textbf{switchr} to reproduce a published
result. Specifically, we recreate a subset of from 
\cite{anders_differential_2010}'s paper presenting methods
implemented in the \textbf{DESeq} package\footnote{With generous permission from the authors.}. The authors
present many results in their paper. We focus here on one: the number
of differentially expressed genes found within their fruit fly data.

The authors' original code will not run under modern versions of
\textbf{DESeq} \citep{becker_gran_2014} due to evolution of the API.
A direct port of the code to the new API yields radically different
results, with approximately half as many genes are identified. Thus, we
cannot assess the strict replicability of their results using modern
versions of the package.

Fortunately, Anders and Huber provided code,
data, and \verb!sessionInfo()! output within their supplementary
materials. With \textbf{switchr} --- and access to the correct
version of \proglang{R} --- the information they provide is sufficient to
reproduce their results. 

Given a compatible version of R, our first step is to reproduce the
authors' software context. \textbf{switchr} provides two ways to \emph{switch to} a compatible
context. The first is to seed  a package library directly with the published
\verb!sessionInfo()! output (console output omitted for brevity
here and below):

\begin{Code}
switchTo( "DESeqRepro", seed = sessInfo)
\end{Code}

The exact package versions listed in
\verb!sessInfo! are retrieved and installed into the
"DESeqRepro" library. This code, however, requires the work
of locating, retrieving, and building the packages to be duplicated on
each machine. This is inefficient if many users will be recreating the
same context; using \textbf{GRANBase}, we can perform these
retrievals once and publish the resulting cohort in a package 
repository via the \verb!makeRepo()! function:

\begin{Code}
makeRepo(name = "10.1186/gb-2010-11-10-r106",
	 manifest = sessInfo,
	 repo_dir = destDir)
\end{Code}
We have uploaded the resulting repository to
\url{http://research-pub.gene.com/gran/10.1186/gb-2010-11-10-r106}, and
we presently switch to it:
\begin{Code}
repo <- "http://research-pub.gene.com/gran/10.1186/gb-2010-11-10-r106"
switchTo("DESeqRepro", seed = repo)
\end{Code}

As noted above, the above code must be run in a compatible version of
R. We achieved this by using a virtual machine, specifically an Amazon
Machine Image (AMI), containing \proglang{R} 2.12 with \textbf{switchr}
installed. The specific AMI we used is freely available here: \href{http://thecloudmarket.com/image/ami-8afc30e2--ubuntu-natty-r-2-12-1}{ami-8afc30e2}.

Having approximately recreated Anders and Huber's original software
context, we are ready to reproduce the result:

\begin{Code}
# (re-)run in R 2.12
suppressMessages(library(DESeq))
countsTableFly <- read.delim( "fly_RNA_counts.tsv" )
condsFly <- c( "A", "A", "B", "B" )
# Add dummy names to avoid confusion later
rownames( countsTableFly ) <- paste( "Gene", 1:nrow(countsTableFly),
				    sep="_" )

cdsFly <- newCountDataSet( countsTableFly, condsFly )
cdsFly <- estimateSizeFactors( cdsFly )
cdsFly <- estimateVarianceFunctions( cdsFly )
resFly <- nbinomTest( cdsFly, "A", "B" )
length( which( resFly$padj < .1 ) )
\end{Code}

\begin{verbatim}
[1] 864
\end{verbatim}

This agrees with the original results. Thus we have successfully
replicated published results which modern software contexts will no
longer produce (with good reason).

\subsection{GitHub-based packages and package manifests}
\label{sec-4-2}

The \textbf{switchr} and \textbf{GRANBase} packages provide two
very different mechanisms for interacting with packages and package
versions not currently
hosted in a repository: direct installation
and the construction of validated repositories which contain such
packages, respectively. We illustrate these approaches below in the context of
Github-based packages.

General package manifests are created using the \verb!Manifest()!
constructor, but for typical GitHub use-cases, the specialized
\verb!GitHubManifest()! constructor is more convenient:

\begin{Code}
ghman = GithubManifest("gmbecker/rpath", "hadley/lazyeval",
    "hadley/dplyr", "rstudio/ggvis")
\end{Code}

 To install the development version of \textbf{ggvis}  --- and its dependencies,
\textbf{lazyeval} and \textbf{dplyr} --- we use the 
\verb!install_packages()! function. We pass our package manifest as the
'repository' from which the packages will be installed (installation
output omitted for brevity).

\begin{Code}
install_packages("ggvis", ghman)
\end{Code}

The above call generates a JIT repository containing the necessary
packages and installs them using standard \proglang{R} machinery in a single
step. 

In this case, the code does not appear substantially different from
calling \verb!install_github()! multiple times; manifests,
however, can be shared, and can list many more packages than those
needed for a particular install.  With a sufficiently exhaustive,
centralized manifest, only the \verb!install_packages()! call is required.

We can install the packages in the manifest
directly by \emph{seeding} a new library with our manifest. By default,
seeding installs all packages listed directly in the manifest, but we
can restrict the installation to a specific set of packages. Furthermore, when
specifying the packages to install directly, we can include packages
which appear in the manifest's dependency repositories (Bioconductor and
CRAN by default) but not in the manifest itself. The \textbf{XML}
package is such a package in the second expression below.

\begin{Code}
switchTo(name = "githubLib", seed = ghman)
switchTo(name = "githubLib2", seed = ghman,
	 pkgs = c("rpath", "XML"))
\end{Code}

While \textbf{switchr} provides direct installation of package
cohorts,  the \textbf{GRANBase} package allows users or
organizations to construct validated repositories from 
package manifests. This involves passing a package
manifest or existing \textit{\textbf{GRANRepository}} object to the
\verb!buildSingleGRANRepo()! function, as shown below. We refer
system administrators to the \textbf{GRANBase} documentation for a
complete discussion of this process and the options available.

\begin{Code}
repo = makeRepo(ghman, cores = 3,
    basedir = tempfile("testrepo"))
\end{Code}

This process generates a standard, validated \proglang{R} package repository,
which can be queried via \verb!available.packages()! and installed
from via \verb!install.packages()! or \textbf{switchr}'s
\verb!install_packages()!. We prefer the latter, because it records
the package provenance.

\begin{Code}
available.packages(repo)
install_packages("ggvis", repo)
\end{Code}

\subsection{Publishing and distributing package manifests}
\label{sec-4-3}

The \verb!libManifest()! function allows users to "reverse seed" a
package library, generating a session or package manifest representing
the packages installed therein. 

\begin{Code}
switchTo("MyProject")
mani <- libManifest()
\end{Code}

For packages installed using \textbf{switchr} --- whether manually or while
seeding a library --- the annotations added to the installed
DESCRIPTION files are used to (re-)construct the manifest entry. For
those installed via other methods, \textbf{switchr} attempts to
locate information about the package automatically. By default,
package names are matched against existing CRAN and Bioconductor
packages. Optionally, users can also provide an existing manifest to be
searched as necessary.

We also support the serialization of a package or seeding manifest as
a tab-delimited plaintext data file by passing a path to a local file
(or connection) to \verb!publishManifest()!. 

\begin{Code}
publishManifest(mani, tempfile("mani"))
\end{Code}

The package information is stored in tabular form within the body of
the file, while the type of manifest and dependency repositories are listed within
comments within the header. In the case of a seeding manifest, an
additional column is added indicating version information.

Our \textbf{switchrGist} extension for \textbf{switchr} provides
a mechanism, built upon \cite{scheidegger_github:_2013}'s \textbf{github} package, which publishes manifests directly to
Github gists when \verb!publishManifest()! is passed a gist target
object (constructed via \verb!Gist()!).

\begin{Code}
publishManifest(mani, Gist())
\end{Code}

The \verb!loadManifest()! function reverses the publish operation,
loading a saved manifest from the file or gist URL.

\section{Discussion}
\label{sec-5}
\subsection{Limitations}
\label{sec-5-1}

Our framework has some limitations. First, the internal
representation of a particular class of \proglang{R} object may change over
package versions. This limits our ability to directly compare results
generated by running the same code under different software contexts
within a single \proglang{R} session. The BiocGenerics package provides the
\texttt{updateObject()} generic for updating object representations. However,
relying on object conversion decreases the probability of a valid
comparison.

Secondly, many old package versions are not installable
within modern versions of R, and older \proglang{R} versions can be difficult to
install themselves. One way to mitigate this issue is to use virtual
machines which provide various historical \proglang{R} versions, as we did in
Section \ref{sec-4-1}. These images would be bare-bones, potentially
providing only \proglang{R} and \textbf{switchr}, and would allow users to re-use images, 
rather than build a complete virtual machine for each analysis, as
proposed by \cite{howe_virtual_2012}. We could also take advantage of
the Bioconductor AMIs, which
correspond to all minor and most micro releases of \proglang{R} (i.e., all
Bioconductor releases) since \proglang{R} 2.8 \citep{bioconductor_team_bioconductor_2014}.
Another approach is to use \textbf{switchr} within the context
of a more stringent reproducibility framework, such as \emph{rocker}
\citep{boettiger_rocker-org/rocker_2014}, which seeks to build \proglang{R} within
Docker \citep{docker_inc_docker_2013} images.  

Finally, installation from a source code repository or archive assumes
the ability to build packages from source.  In cases where this is
difficult or otherwise infeasible, the offending packages can be
pre-compiled and placed in a package repository.

\subsection{Future directions}
\label{sec-5-2}
\subsubsection{Generalizing the base R installation mechanism}
\label{sec-5-2-1}

From both the technical and conceptual standpoints, our manifest-based
installation framework is similar to the one included with R. The
PACKAGES(.tgz) file in a package repository is a combination of a
centralized package manifest and a cache of package dependency
information, tightly coupled with a cohort of pre-built packages.

The core installation mechanism of \proglang{R} could be extended to include some or all
of the concepts discussed here --- decentralized repositories,
versioned installation, direct installation from SCM --- with only
narrow, targeted changes to existing code. For example, one could add
a new field to the PACKAGES file that indicates the location of the
package source code and avoid the assumption that the
source tarball is present on the same host. This would enable
anyone with access to Github or other
web host to define package cohorts and would help democratize
scientific software publication.

\subsubsection{Mapping seeding manifests to DOIs}
\label{sec-5-2-2}

Modern publications are typically referenced by \emph{digital object
identifiers} (DOIs). With a DOI, one can lookup a publication or its
metadata, such as its citations \cite{pila_crossref.org_2013}.
A simple, centralized repository mapping DOIs to recreatable software contexts
would dramatically increase the strict reproducibility of published
results for which the code and data are available. The authors might
simply upload their shared software context as a Gist, or make the investment
of deploying a centralized package repository, which streamlines
installation. Recreating the
software context for a particular publication, then, could be as easy
as \emph{switching to} a URL, e.g., 

\begin{Code}
switchTo("repro", seed = "http://Rlibraries.org/<doi>")
\end{Code}

\subsubsection{Dynamic documents and reproducibility}
\label{sec-5-2-3}

\cite{gentleman_statistical_2004} argue that for true reproducibility, a dynamic
document should be distributed within a \emph{compendium} that also
includes the data and software required to run the document. With the ability of \textbf{switchr}
to recreate historical contexts from
package archives and SCM systems, encoding a \verb!sessionInfo()!
object, or a seeding manifest, within a dynamic document is essentially
equivalent to explicitly including the package sources themselves (the
approach pursued by 
\cite{rstudio_inc_packrat_2014} with their \textbf{packrat}
system). Gehring and Becker have preliminary, unpublished
work in this area which suggests that this encoding can be fully
automated within the context of re-runnable dynamic
documents. 

We also imagine dynamic documents which contain code chunks
intended to be run using different package libraries or even R
versions. One example of this would be an automated report describing
the differences in results when running the same code within different
contexts, such as our \textbf{DESeq}-based example in Section \texttt{DESeqApp}. The
\textbf{switchr} framework, possibly combined with virtual 
machines, such as the AMIs provided by the \cite{bioconductor_team_bioconductor_2014}, gives
dynamic document processing systems the flexibility necessary to
support this use-case automatically.

\subsubsection{Approximating software contexts based on historical CRAN states}
\label{sec-5-2-4}

As we saw in Section \ref{sec-2}, \textbf{switchr} enables us
to proactively generate a virtual snapshot of the current state of a
repository and to materialize the snapshot on a later date. We can
also generate snapshots retroactively.

The \textbf{crandb} package and database
\citep{csardi_crandb:_2014} is a prototype system that automatically
retains extensive, queryable information about the contents of CRAN.
\textbf{switchr} accesses \textbf{crandb} to enable users
to install cohorts of packages
corresponding to historical dates, releases of \proglang{R}, or releases of
specific packages. This can provide approximate recreation of
\proglang{R}-based software contexts for publications which do not provide
version information in the form of a seeding manifest or
\verb!sessionInfo()! output. Bioconductor could track and provide
similar data, further extending the utility of this approach.

\section{Availability}
\label{sec-6}

We have released our \textbf{switchr}, \textbf{switchrGist}, and \textbf{GRANBase}
packages under the Artistic 2.0 open-source software license.  Current,
up-to-date development versions of their source-code is available on GitHub at
\url{https://github.com/gmbecker/switchr},
\url{https://github.com/gmbecker/switchrGist}, and
\url{https://github.com/gmbecker/gRAN}, respectively. Stable, release
versions of the packages can be found on CRAN (XXX TODO)

\bibliography{GRANPaper}
\end{document}